\documentclass[sn-mathphys-num]{sn-jnl}

\usepackage{graphicx}%
\usepackage{multirow}%
\usepackage{amsmath,amssymb,amsfonts}%
\usepackage{amsthm}%
\usepackage{mathrsfs}%
\usepackage[title]{appendix}%
\usepackage{xcolor}%
\usepackage{textcomp}%
\usepackage{manyfoot}%
\usepackage{booktabs}%
\usepackage{algorithm}%
\usepackage{algorithmicx}%
\usepackage{algpseudocode}%
\usepackage{listings}%

\begin{document}

\title[Deep Learning-Based Beamlet Model]{Deep Learning-Based Beamlet Model for Generic X-Ray Beam Dose Calculation} 


\author[1]{\fnm{Maxime} \sur{Rousselot}}\email{rousselot.maxime@gmail.com}

\author[1]{\fnm{Jing} \sur{Zhang}}\email{jing.zhang@univ-brest.fr}

\author[1]{\fnm{Didier} \sur{Benoit}}\email{didier.benoit@inserm.fr}

\author[1]{\fnm{Chi-Hieu} \sur{Pham}}\email{chi-hieu.pham@univ-brest.fr}

\author*[1]{\fnm{Julien} \sur{Bert}}\email{julien.bert@univ-brest.fr}

\affil*[1]{\orgdiv{LaTIM}, \orgname{INSERM UMR1101, University of Brest}, \orgaddress{\street{Camille Desmoulins}, \city{Brest}, \postcode{29238}, \country{France}}}


\abstract{\textbf{Purpose:} Modeling the absorbed dose during X-ray imaging is essential for optimizing radiation exposure. Monte Carlo simulations (MCS) are the gold standard for precise 3D dose estimation but require significant computation time. Deep learning offers faster dose prediction but often lacks generality, as models are typically trained for specific anatomical sites and beam geometries. The aim in this work was proposing a generic deep-learning approach for dose calculation that can be used for multiple X-ray imaging systems.

\textbf{Methods:} This article proposes a versatile approach combining beamlet decomposition with deep learning, where the X-ray beam is broken down into beamlets. By using a sampling approach, various beam shapes can be generated, reducing learning complexity. The model learns the dose response of a beamlet for different energies and patient properties, making it adaptable to new system geometries without altering the learning model. In this work, we propose combining two U-Net networks (1D+3D) trained on different body parts to predict the dose of a beamlet regardless of its orientation and energy. 

\textbf{Results:} Results have shown that the deep learning-based dose engine achieved a relative dose error of approximately 1.2$\pm$3.87\% compared to the reference dose. For a more realistic simulation in cone-beam CT, dose results exhibited a relative error within the beam of 5\% compared to a full MCS. The convergence of the proposed method was faster compared to MCS, with a speedup of 130 times for equivalent dose results.

\textbf{Conclusion:} In this work, a hybrid method combining beamlet random sampling and deep learning approaches was explored. The versatility of the proposed solution allows for the simulation of multiple X-ray systems without the need to retrain the deep learning model with new beam specificities. The same trained model is capable of calculating the 3D dose within the patient for helical CT, cone-beam CT, fan-beam CT, or any collimated beam shape.
}




\maketitle


\section{Introduction}\label{sec1}
Modeling the absorbed dose delivered to a patient during an X-ray medical imaging procedure is an essential tool for optimizing, monitoring and reducing exposure to ionizing radiation. In conventional radiology, the Dose-Area Product (DAP) is used to calculate the dose to the skin. However, for certain applications, such as pediatric imaging or protocol optimization, it is crucial to estimate the specific dose to the patient by considering their anatomy, as well as in 3D at the organ level. Within this context Monte-Carlo simulations (MCS) are considered to be the gold standard to estimate this deposited dose. Especially when the absorbed dose need to be estimated in 3D within the organs. They are considered more accurate than any other traditional deterministic dose calculation engine such as pencil beam (PB)~\cite{mohan1986differential} or collapsed cones(CC)~\cite{ahnesjo1989collapsed}, etc. MCS provide more precision around tissue heterogeneities and air cavities~\cite{Krieger_2005}. Indeed, they model the comportment of numerous particles, taking into account density variations within the patient with the help of CT images and simulate particle transportation with high precision and score energy depositions accordingly. However, to enable sufficient statistical accuracy, the trajectories and physical processes of millions of particles have to be calculated through the patient voxelized phantom. This requires a considerable amount of execution time. To reduce the computation burden, several approaches were proposed. For example, as MCS are highly parallelizable, GPU implementations offer sensible gain in terms of speed~\cite{Bert_2013}~\cite{tian_gpu_2015}. Other methods, such variance reduction techniques (VRT), proposed to modify the calculation of particles histories to increase the efficiency of MCS without introducing approximation. For example, the sampling of the particle tracking can be optimized~\cite{Behlouli_2018}, or the photon dose deposition along its path can be condensed with the track length estimator (TLE) method ~\cite{Williamson}.

In recent years, the fast evolution of deep learning opens a new way to quickly and accurately predict a dose distribution in various applications. Many specialised neural network were train to predict the dose to optimize particle therapy, for example, for helical tomotherapy~\cite{Liu}, intensity-modulated radiation therapy (IMRT)~\cite{kontaxis2020deepdose}, brachytherapy~\cite{Mateo2021}, proton therapy~\cite{neishabouri2021long}. Deep learning approach were also widely used for medical imaging such as positron emission tomography (PET)~\cite{lee2019deep}, single photon emission computed tomography (SPECT)~\cite{G_tz_2020}, and for x-ray imaging, CT scan~\cite{Roser} and Cone Beam CT \cite{villa_fast_2023}. 

In X-ray imaging, a significant limitation of deep learning approaches is their lack of generality. Typically, the network model is trained for a specific anatomical site (e.g., pelvis, head, thorax) and beam geometry (e.g., cone beam, fan beam). This specialization makes it challenging to parameterize the beam with varying aperture angles or collimation settings. One solution, proposed by \cite{villa_fast_2023} for thoracic X-ray imaging, involves using a conditional network with parameters such as beam position, angulation, and X-ray tube voltage. However, this approach did not address collimation due to the combinatorial complexity of multiple beam geometries. In radiotherapy, where beam collimation is crucial, \cite{kontaxis2020deepdose} suggested encoding the beam into an image via raytracing as input to the network. This method requires extensive pre-processing and large training datasets to cover all possible collimation variations, making the learning process time-consuming and necessitating a more complex network. Consequently, each trained network is often limited to a specific application and lacks adaptability.

In this article, we present a more versatile approach that integrates beamlet decomposition with deep learning. The core concept involves breaking down the X-ray beam into elementary components called beamlets, similar to the method used in radiation therapy inverse planning \cite{unkelbach_optimization_2015}. A beamlet is a pencil beam that serves as a fundamental element of any X-ray beam. By employing a sampling approach, various beam shapes, such as fan beam, cone beam, and collimated beam, can be randomly generated from these beamlets. This reduces learning complexity, as the beam shape and system configuration are determined through sampling rather than by the network itself. The model only needs to learn the dose response of a beamlet for different energies, orientations, and patient anatomical properties. This approach is more generic because adjusting the sampling method can accommodate new system geometries, while the learning model remains unchanged as it depends solely on the patient.

\section{Materials and methods}

\subsection{Database creation}  
\begin{figure}[ht]
	\includegraphics[width=\textwidth]{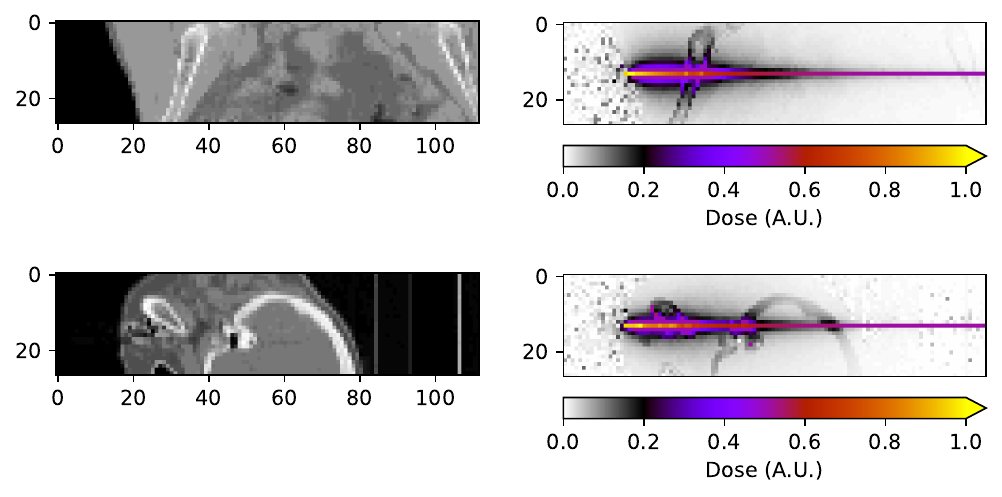}
	\caption{Examples of 3D patches for a photon beamlet of $89$ keV (first row) and a photon beamlet of $428$ keV (second row). The left column is the central slice of each CT image patch and the right column the corresponding absorbed dose map in normalized arbitrary unit estimated by Monte Carlo simulation.}
	\label{fig:Patchexample}
\end{figure}

For the training data set a total of 80 CT patients from different anatomical site (head and neck, thorax and pelvis) was used. These retrospective data were collected and anonymized by the Brest University Hospital. All images were re-scaled to a voxel resolution of $3\times3\times3~mm$. A total of 17000 beamlets were generated with random position, orientation, energy and patient. For each of these beamlets a 3D CT image patch from the corresponding patient were extracted following the direction of the beamlet. Then the oriented patch was transformed into axis aligned image, in order to be interpretable by the network. Each patch corresponds to a small region of the patient CT where the beamlet is passing through, and will be used to recovered the absorbed dose. A patch-size of $27\times27\times112$~voxels was defined. Since the beamlet is a pencil beam, the patch size along the particle direction was longer. Those dimensions are a trade-off between optimal data size and the relevance of the deposited dose gradient considering a maximum photon energy of $<1$ MeV. The absorbed dose within each patch was calculated using Monte Carlo simulations with the GATE software~\cite{SarrutGate}. Each patch, that contains Hounsfield value, was converted in materials labels using density conversion given by ~\cite{Schneider_2000}. A total of $5\times10^8$ particles was run for each beamlet in order to reach for every voxel a dose statistical uncertainty less than 5\%. Some examples of extracted patches and the resulting dose can be found in Figure~\ref{fig:Patchexample}.

\subsubsection{Patch extraction}
In order to limit the size of the patch and the network, only pixels around the beamlet are considered in the dose prediction. As the beamlets are oriented, it is necessary to be able to extract the patch from the CT image based on the beamlet's orientation. This is mandatory since the network architecture used only axis-aligned image. Therefore, the oriented patch image $\mathbf{I}_{op}$ has to be transformed into an image axis-aligned patch $\mathbf{I}_{aap}$. Similarly, the axis-aligned dose patch after prediction has to be transformed and placed back into the dose image at the right beamlet position.

This was done by first calculating the entry point of the beamlet into the patient's CT image. To achieve this, the beamlet is treated as a parametric line (ray) and defined as follow,

\begin{equation}
	\mathbf{b}=\mathbf{b}_o + \mathbf{d}l 
\end{equation}

with $\mathbf{b}=\{b_x, b_y, b_z\}^T$ a point along the beamlet path $(L)$, $\mathbf{b}_o=\{b_{ox}, b_{oy}, b_{oz}\}^T$ the origin position of the source, $\mathbf{d}=\{d_x, d_y, d_z\}^T$ the direction of the beamlet, and $l$ the parametric distance along the line $(L)$. CT images are naturally axis-aligned images. Then the bounding geometry of the CT phantom is defined as an Axis Aligned Bounding Box (AABB). AABB objects are aligned with the axis $\{u_x, u_y, u_z\}$ of the simulation. Therefore, the intersection between the beamlet and the CT image was solved using an efficient ray/box intersection algorithm \cite{smits_efficient_2002}. The interaction point between the beamlet and the AABB was determined by considering each intersection of the beamlet with the slabs that compose the AABB. A slab is defined as the surface between a pair of parallel planes or lines. Every distance between the ray and the minimum and maximum boundary slabs were calculated using their respective plane equations. The final intersection distance $l_\Box$ with the AABB, was given by the minimum positive value between all slabs intersections. The point $\mathbf{p}_{\Box}$ that intersect the phantom was subsequently calculated using the parametric line equation.

Based on this intersection point and the beamlet direction an affine transformation matrix $\mathbf{M}$, that transform the beamlet oriented patch to be extracted $\mathbf{I}_{op}$ in an axis-aligned patch $\mathbf{I}_{aap}$, was defined. This matrix was composed of a $3\times3$ sub-matrix for the rotation $\mathbf{R}_{xyz}$ and a $1\times3$ sub-matrix for the translation $\mathbf{T}_{xyz}$, as follow,

\begin{equation}
	\mathbf{M} =
	\left\{
	\begin{array}{cc}
		\mathbf{R} & \mathbf{T} \\
		\mathbf{0} & 1
	\end{array}
	\right\}
\end{equation}

A backward sampling approach was used to align the patch. This avoids aliasing and holes in the images (voxel without values) due to numerical approximation. Instead of transforming directly the oriented patch image into axis-aligned $\mathbf{I}_{op} \stackrel{\mathbf{M}}{\longrightarrow} \mathbf{I}_{aap}$ with the matrix $\mathbf{M}$, the inverse process was used. For each pixel of the targeted image $\mathbf{I}_{aap}$, the inverse matrix $\mathbf{M}^{-1}$ was used to find the corresponding voxel position in $\mathbf{I}_{op}$. The value of this voxel was then copied into the targeted image $\mathbf{I}_{aap}$. After predicting the dose of the current CT patch, the corresponding dose is placed back, by following the same backward sampling process by directly using the transformation $\mathbf{M}$. In practice, to ensure extremely fast execution of patch extraction, the voxel values are not copied. Instead, memory pointers are stored, creating a direct link between the original image and the aligned one. Additionally, parallel programming was implemented so that each CPU thread manages a portion of the patch, thereby improving efficiency.

\subsection{Neural networks architecture}
\begin{figure}[htbp]
	\centering
	\includegraphics[width=0.7\textwidth]{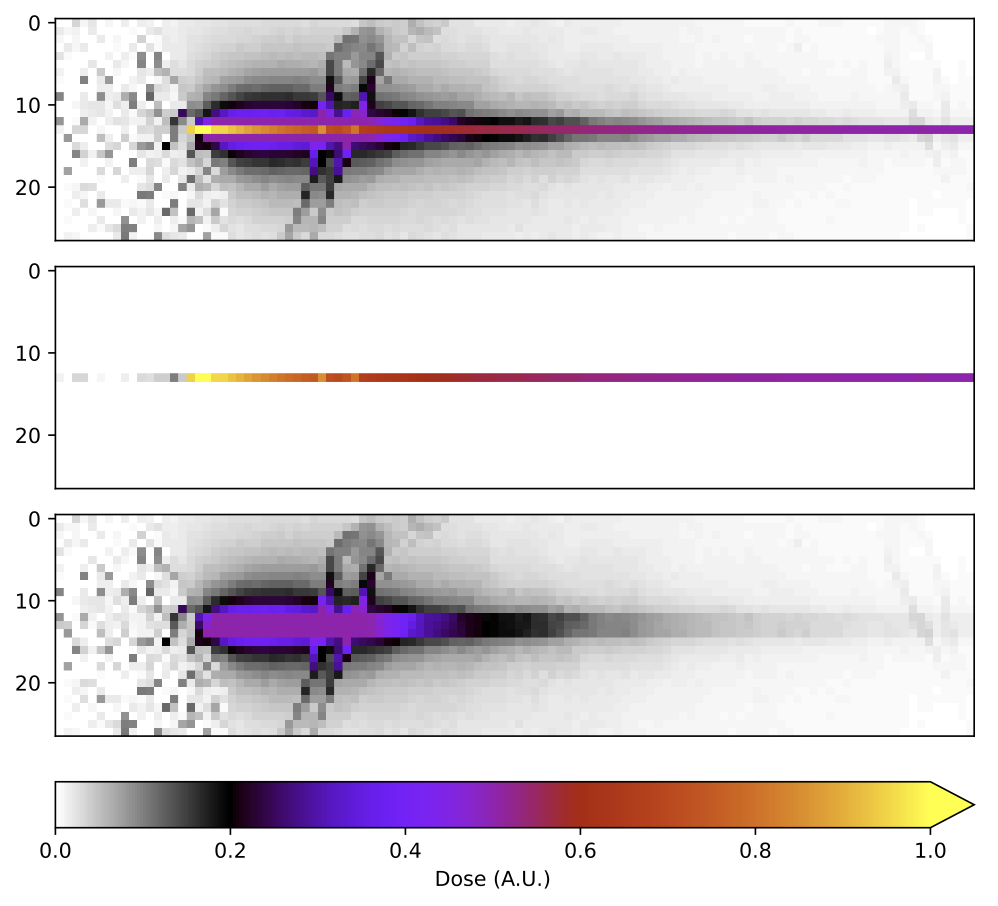} 
	\caption{A full dose map from a beamlet Monte Carlo simulation (top) can be decomposed into primary dose deposition (middle) and scattered dose deposition (bottom).}
	\label{fig:beamlet}
\end{figure}

For predicting the dose corresponding to each patch and each beamlet energy, a Convolutional Neural Network (CNN) was designed. The main idea was to have an architecture with two inputs: the CT patch and the energy of the beamlet. The output of the network was the absorbed dose map. However, as emphasized in Figure~\ref{fig:Patchexample}, doses have two distinct areas of scale values. The first area has high dose values along the beamlet direction, where primary interactions occur. The second area has smaller dose values around the beamlet, corresponding to particles scattered within the patch. Between primary and scattered dose values, a magnitude difference of two orders was measured. Since CNNs are generally less efficient at learning high frequencies, we chose to split the dose map (Figure~\ref{fig:beamlet}) into two dose maps: the dose from the primary interaction and the dose from scattering.

\begin{figure}[ht]
	\includegraphics[width=\textwidth]{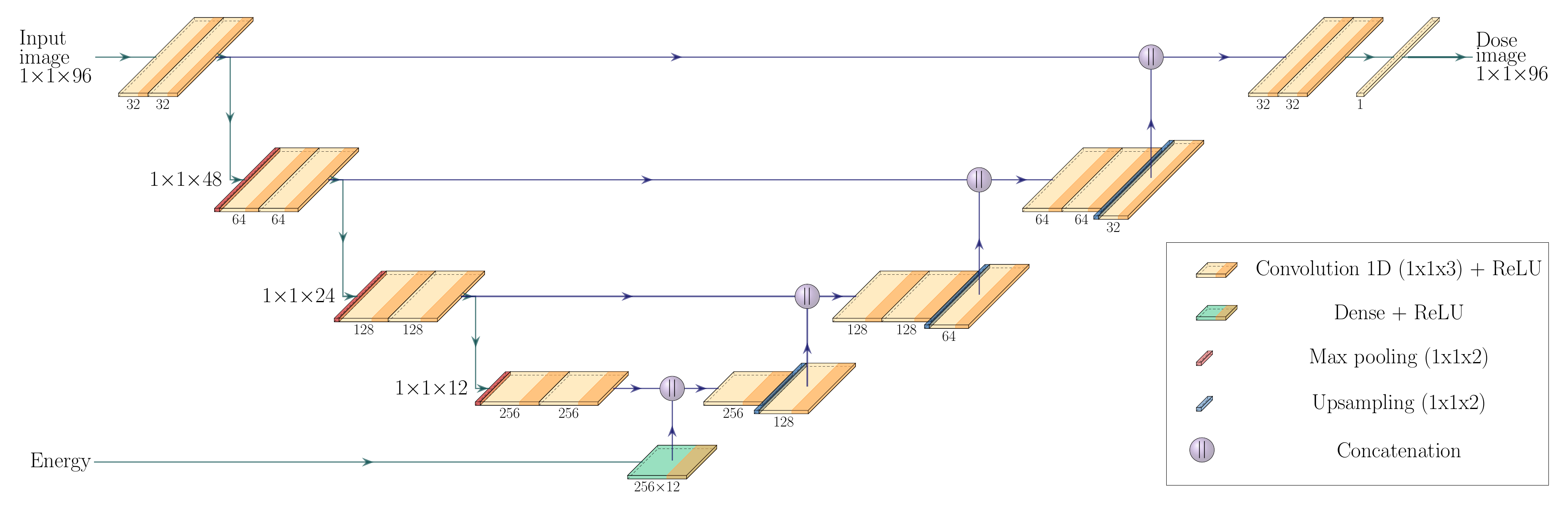}
	\caption{First 1D network that predict the dose of the beamlet primary interactions.}
	\label{fig:CNN1D}
\end{figure}

Two sub-networks were then designed for predicting the dose for each beamlet and patch. Both were based on U-Net~\cite{u-net_2015} architecture, which was already proven to be efficient for dosimetric application \cite{villa_fast_2023}. We do not use more complex architectures, such as Transformers, because the aim is to have a lightweight network that can predict the dose as quickly as possible. That kind of well-know network is composed of one encoder and one decoder, which gives the network its U-shape architecture. We include the energy input in the latent space using a dense layer. The aim was conditioning the network considering the beamlet energy. The energy value was not injected with the image patch at the input of the network, because during the training the single energy value will be mostly discarded compared to the thousand of voxels that compose the patch image.
The first network (Figure~\ref{fig:CNN1D}) aims to predict the dose of the primary interactions. Then, has it was a beamlet, the input and output images, as well as all layers filter, are uni-dimensional. We used a typical dimension for the different layer filters: 3 for the convolution filters and 2 for the max pooling and upsampling layers.

\begin{figure}[ht]
	\includegraphics[width=\textwidth]{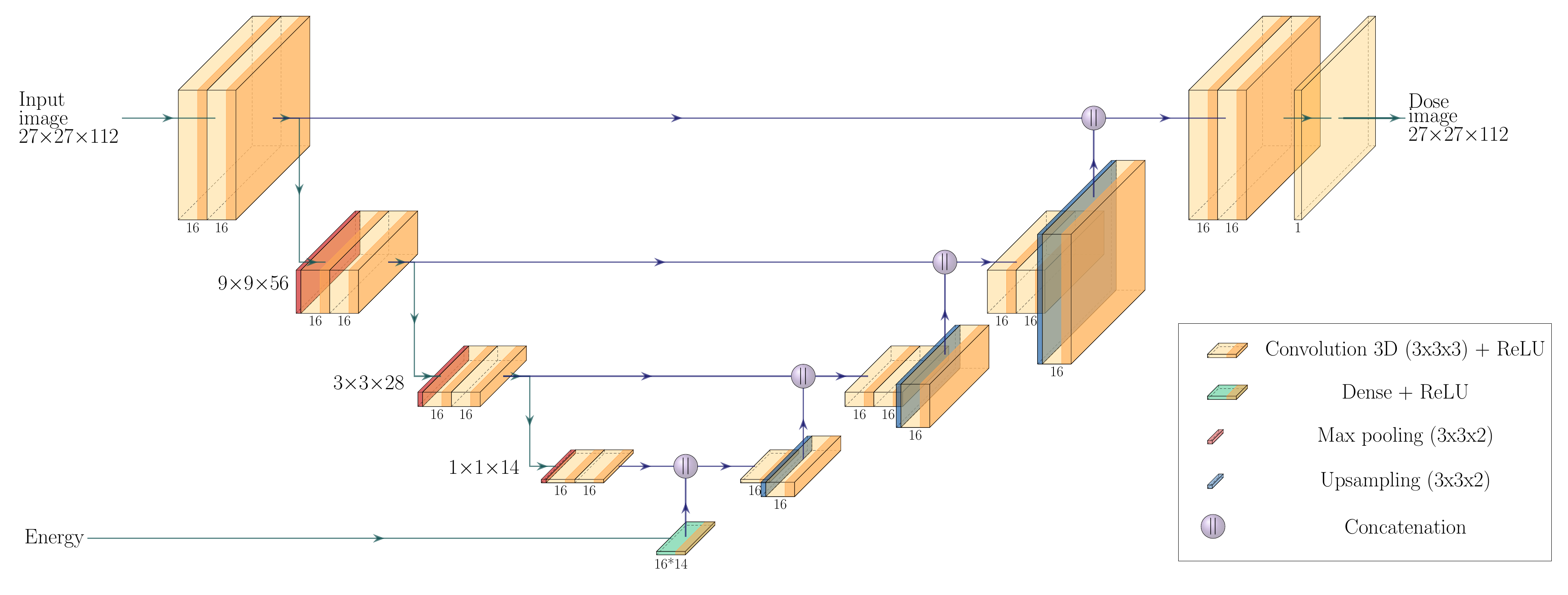}
	\caption{Second 3D network that predict the dose of the beamlet scattered interactions.}
	\label{fig:CNN3D}
\end{figure}

The second network that predict the 3D dose maps of the scattered particles is shown in Figure~\ref{fig:CNN3D}. Since the patch is in 3D, the convolution filters have also to be three-dimensional with a size of $3\times3\times3$. Moreover, to preserve the symmetry of the dose across the network, a maximum pooling and upsampling filter of size $3\times3\times2$ instead of the more conventional size $2\times2\times2$ was chosen. 

\subsection{Training and validation}
The dataset was divided into three sets, the training, the validation, and the test set, by randomly selecting, 13500, 1500, and 2000 patches, respectively. For the test set patches, different patients, which was not used for the training and the validation was used. The aim was to ensure that our network can generalize well on new data. To train the 3D CNN, a classical data augmentation technique was used. We applied three $\mbox{90}^\circ$ rotations along the beamlet axis on the input and output images to quadruple the size of the training set. The dose values were normalized between 0 and 1 to facilitate the training. In addition, the absorbed dose to voxel that contains air material (very small density) was set to zero avoiding too much attention to the network on very high dose value that are not relevant for patient dosimetry.

To train the two networks, the adaptive momentum algorithm (ADAM) ~\cite{kingma2014adam} was used to minimize the mean absolute error between the predicted dose and the MCS doses. During the training, both the validation and training loss at each epoch were monitored. The process was stopped when the validation loss ceased to decrease compared to the training loss. The two networks were implemented using Keras and Tensorflow on an Nvidia GeForce RTX 1080Ti GPU and with a processor Intel(R) Xeon(R) CPU E5-2680 v4 @ 2.40GHz.

\subsubsection{Beam geometry sampling}
The second element of our method was the construction of the X-ray beam geometry using beamlets. When the beam geometry can be analytically modeled, sampling its shape with beamlets becomes straightforward. Beamlets can simply be generated randomly and uniformly, following the origin, direction, and energy distribution of the source spectrum. For instance, in the case of a point source, the beamlet direction is sampled according to isotropic emission angles, i.e., ${\theta, \psi} \in 4\pi$. For a cone beam, the angles are sampled within ${\theta, \psi} \in \left[-\frac{\alpha}{2}, +\frac{\alpha}{2}\right]$, and for a fan beam, the sampling reduces to ${0, \psi} \in \left[-\frac{\alpha}{2}, +\frac{\alpha}{2}\right]$, where $\alpha$ is the aperture angle. For collimated sources, whether simple or complex, a rejection method combined with ray tracing can be used. Here, the beamlet is represented as a parametric line, and its intersection with the collimator is computed. Only beamlets that do not intersect the collimator are retained. This approach allows for the simulation of various collimator geometries, such as square collimators used in fluoroscopy or multi-leaf collimators in radiotherapy. If necessary, the origin of the beamlet (i.e., its spatial position) can also be randomly generated to simulate a moving source. Regarding beamlet energy, the source spectrum must be respected. To randomly sample beamlet energy according to specific probabilities, a uniform random process applied to the cumulative density function (CDF) derived from the spectrum can be used.

\subsection{Evaluation study}

\subsubsection{Beamlet dose prediction}
The proposed deep learning approach was first evaluated at the beamlet level. The aim was estimated the level of accuracy of the CNN dose model. For each beamlet containts in the test set the corresponding CT image patch was extracted and used to estimate the dose with our networks. The predicted doses were compared against the beamlet dose from the Monte Carlo simulation consider here as  gold standard. The proposed approach requires two inferences, one to recover the primary dose contribution and another one to get the dose from the scattering. Since absorbed doses are cumulative, the two dose maps were then sum up to obtain the final dose map that will be used for comparison.

Two metrics, the Absorbed Dose Error (ADE) and the Mean Absolute Error (MAE) were calculated in order to estimate the robustness of the networks. The ADE correspond to the total dose absorbed by each patch and was calculated as:

\begin{equation}
	ADE=\frac{|\sum\limits_{i=0}^{N} D_i - \sum\limits_{i=0}^N \tilde{D}_i|}{\sum\limits_{i=0}^{N} D_i}
\end{equation}

where $D_i$ and $\tilde{D}_i$ are respectively the dose at voxel $i$ by the gold standard Monte Carlo simulation and by the CNN. $N$ is the number of voxels that compose the patch. The MAE was calculated as follows:

\begin{equation}
	MAE=\frac{1}{N}\sum\limits_{i=0}^{N} |D_i - \tilde{D}_i|
\end{equation}

\subsubsection{Cone Beam CT imaging}
The core of the proposed method was to combine a sampling beamlet approach with a deep learning-based dose engine to enhance versatility. In Monte Carlo simulations, a cone beam CT is generated by randomly emitting particles that follow the beam geometry, such as the aperture angle or beam direction. The proposed method follows a similar philosophy. Instead of emitting particles, beamlets that mimic the characteristics of any desired complex beam are randomly generated. The absorbed dose of each beamlet that reaches the patient is then estimated using a deep learning method. Similar to Monte Carlo simulations, by sampling enough beamlets, the accumulated dose will converge to a dose map that reveals the dose distribution of the complex X-ray beam. However, the proposed method converges faster because each beamlet simulates a large number of particles predicted by the network, without the need for a full Monte Carlo simulation.

\begin{figure}[ht]
	\centering
	\includegraphics[width=.8\textwidth]{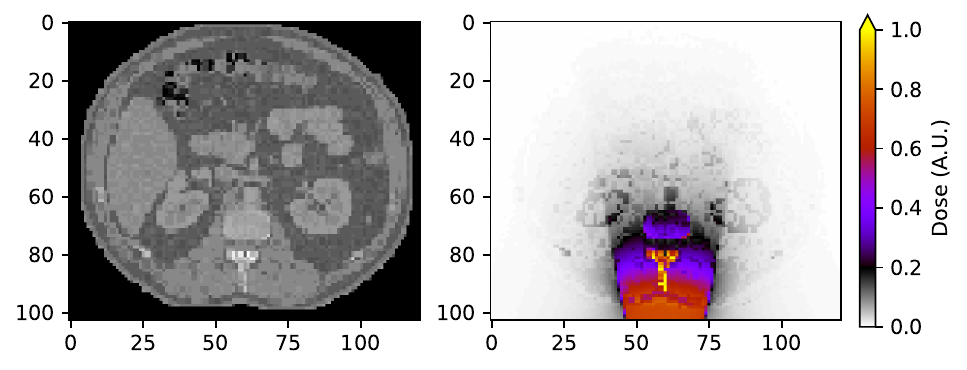}
	\caption{Cone Beam CT imaging application, with a slice showing the patient CT image (left) and the associated absorbed dose map of the beam calculated by MCS (right)}
	\label{fig:GT_ex1}
\end{figure}

To illustrate and evaluate the proposed mechanism, a cone beam CT simulation from a fluoroscopic system, typically used in interventional radiology, was performed. Specifically, an X-ray tube with a rectangular source of $0.5 \times 0.3~mm^{2}$ and an aperture of $10^{\circ}$ was considered. A classical tube voltage of $125$ kVp and a $2$ mm aluminum filter were defined. The X-ray spectrum was obtained using the TASMIP model~\cite{boone1997accurate}. For this simulation, a CT scan of the abdominal region, not included in the training dataset, was used. The CT scan was resized to a resolution of $3\times3\times3~mm^3$ and a size of $123\times81\times80$~voxels. The X-ray source was placed 42 cm from the patient's back. The ground truth was obtained using Monte Carlo Simulation (MCS) with $8\times10^{7}$ particles (see Figure~\ref{fig:GT_ex1}).

Beamlets were randomly generated using the cone beam specifications in terms of energy and geometry. The entry point of each beamlet within the patient was calculated using a ray casting approach. Based on this position and the beamlet direction, the corresponding image patch was extracted on-the-fly. Subsequently, the patch and the beamlet energy were fed into the network, and the resulting dose map was integrated into the final 3D dose map by adding the dose values. This final dose map has the same dimensions as the patient's CT image. The evaluation was performed by comparing the Monte Carlo Simulation (MCS) with the proposed method for different numbers of beamlets. Each time, the Mean Absolute Error (MAE) and Average Dose Error (ADE) metrics were calculated, considering both the dose within the X-ray primary beam, where values are higher, and the dose outside the beam, where scattered doses are smaller.

\section{Results}

\subsection{Beamlet dose prediction}

\begin{figure}[ht]
	\centering
	\includegraphics[width=\textwidth]{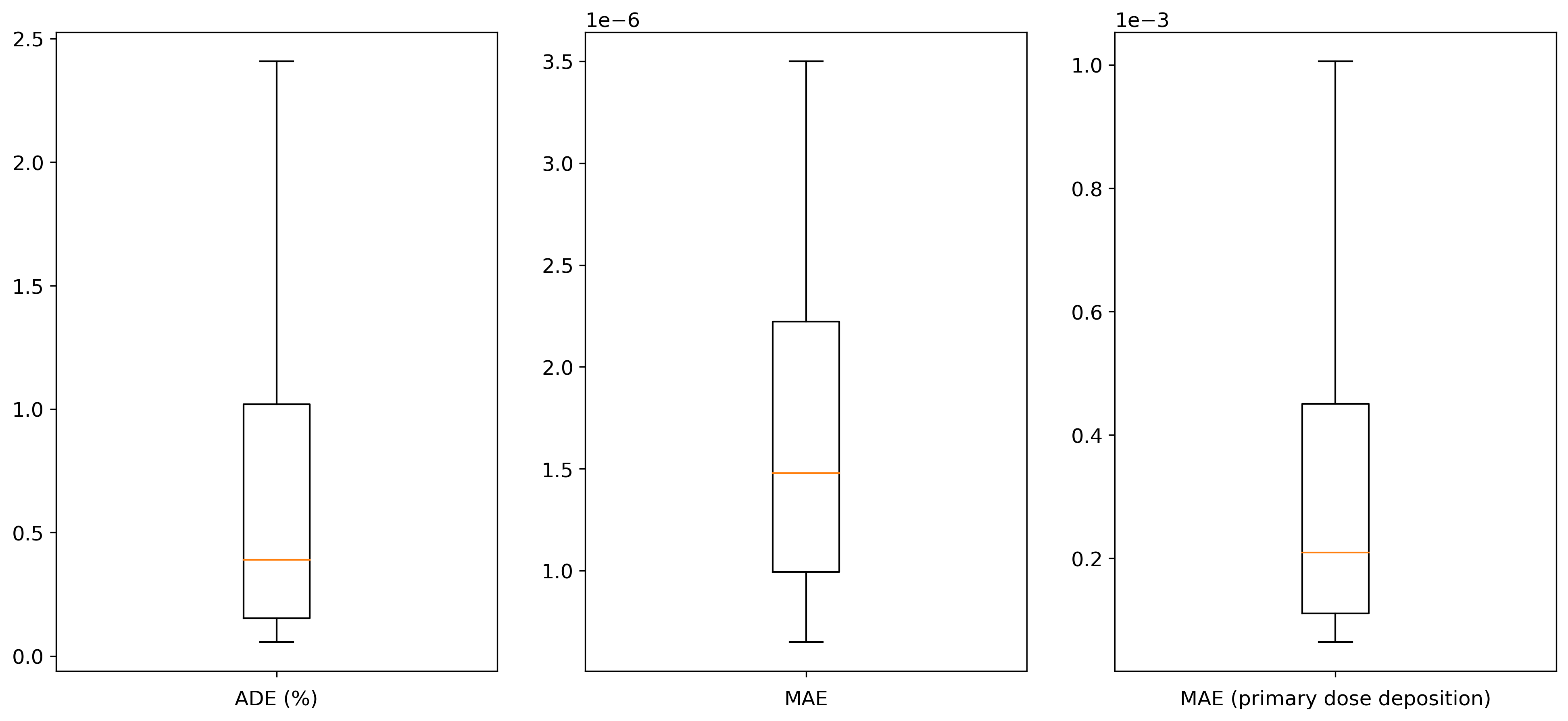}
	\caption{Distribution of the ADE, MAE and MAE only within the primary beam, for the 2000 patches that compose the test set.}
	\label{fig:MAEADE}
\end{figure}

The first network was trained for approximately 4 hours, and the second network for 22 hours. The distributions of the Average Dose Error (ADE), Mean Absolute Error (MAE), and MAE within the primary beam for the test set are shown in Figure~\ref{fig:MAEADE}. Across the 2000 patches that compose the test set, the ADE was $1.2\pm3.87\%$, $1.90\pm1.73\times10^{-6}$ for the MAE calculated on whole patches and $4.6\pm9.4\times10^{-4}$ for the MAE calculated only on the primary dose deposition maps.

\begin{figure}[ht]
	\centering
	\includegraphics[width=\textwidth]{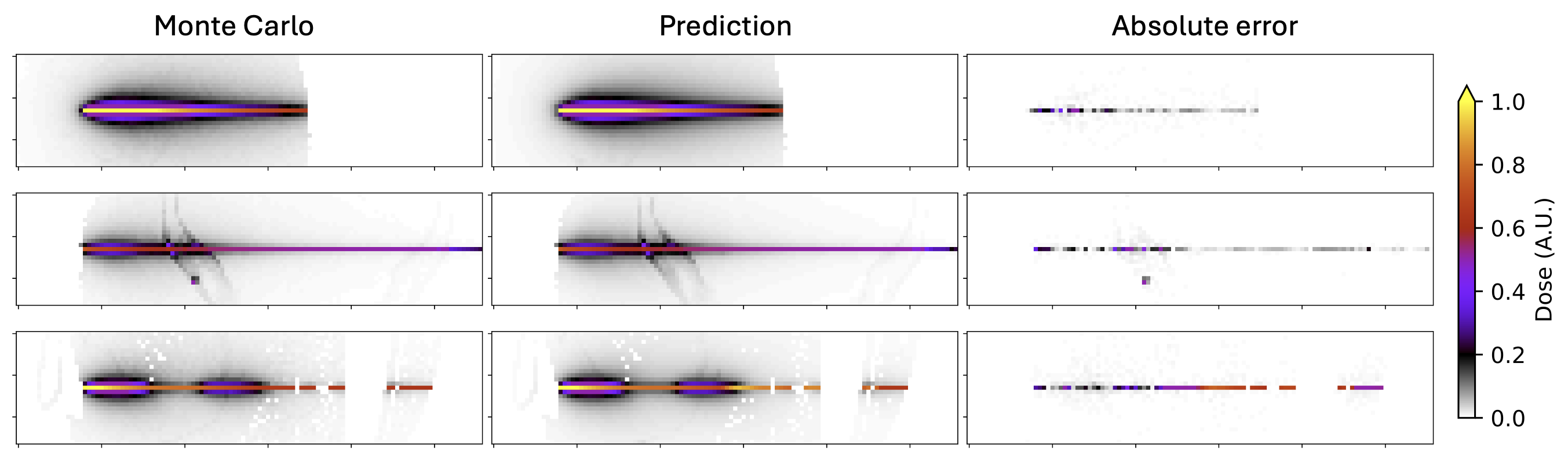}
	\caption{Three examples of predicted dose maps are shown: at the top, the best case with an energy beamlet of 141 keV; in the middle, a median case with an energy of 287 keV; and at the bottom, the worst case with an energy of 287 keV. Each example includes the ground truth from MCS (left), the prediction from the network (center), and the absolute error map (right).}
	\label{fig:monopred}
\end{figure}

Overall, the proposed networks has provided satisfying dose prediction map. Three examples of predicted dose map are shown in Figure~\ref{fig:monopred}, the best and worst case as well as a median case. For the three patches, the ground truth from MCS, the prediction from the network, and the absolute error map were provided. Even in the worst case, the prediction was mainly in a good agreement.

\subsection{Cone Beam CT imaging}
\begin{figure}[ht]
	\includegraphics[width=\textwidth]{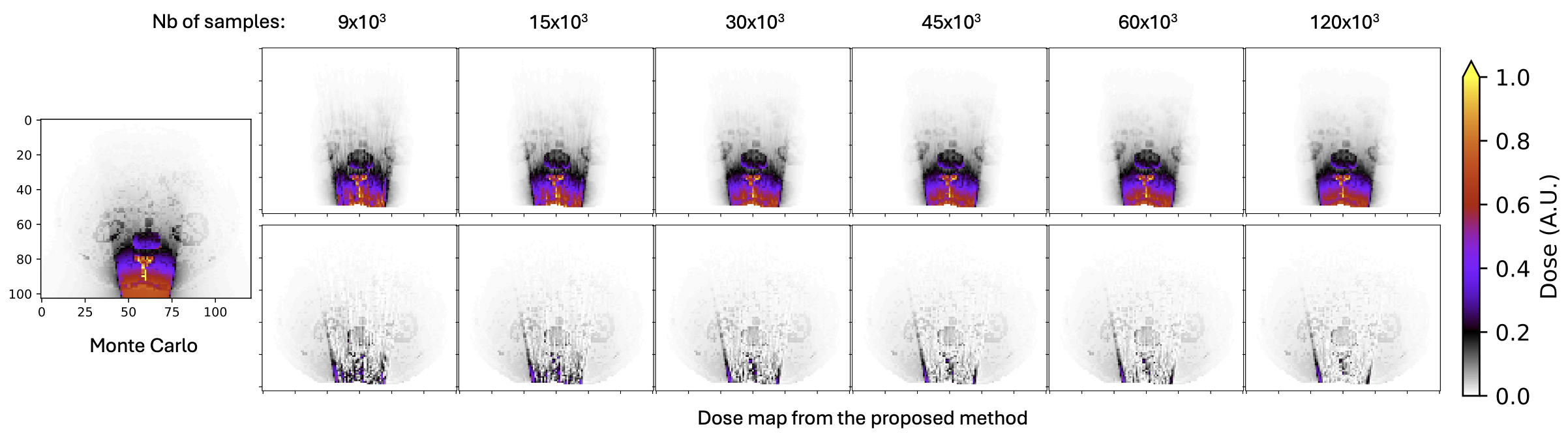}
	\caption{Example of transversal slices for the dose map prediction of the proposed method (first row), and its absolute error (second row) compare to the MCS ground truth (left image).}
	\label{fig:X-RayConeBeam}
\end{figure}

The performance of the proposed method was studied according the number of sampled beamlets. The results are presented in Figure~\ref{fig:X-RayConeBeam}. For each predicted dose map a comparison, using absolute error, was made with the ground thruh provide by the MCS. The first notable observation is that with a relatively small number of beamlets (9,000), we already obtain a fairly comprehensive dose map. As the number of beamlets increases, we quickly converge with 45,000 beamlets to a noise level equivalent to our Monte Carlo (MC) simulation. Beyond this point, a plateau is reached where additional beamlets contribute only marginal improvements. In this evaluation, we identify two limitations to our method. The first is that the out-of-field dose related to scattering in tissues is not fully captured. This issue arises from the patch size. A larger patch would allow for better consideration of doses far from the beamlet's central axis but would also increase prediction time. However, this underestimation of out-of-field dose is not critical, particularly in medical imaging, where the doses involved are extremely low, typically two orders of magnitude smaller than the beam dose. Another limitation highlighted in our evaluation is the known difficulty for the network to predict images at high frequencies. This manifests as a slight error at the beam's border compared to Monte Carlo simulations.

\begin{figure}[ht]
	\centering
	\includegraphics[width=0.85\textwidth]{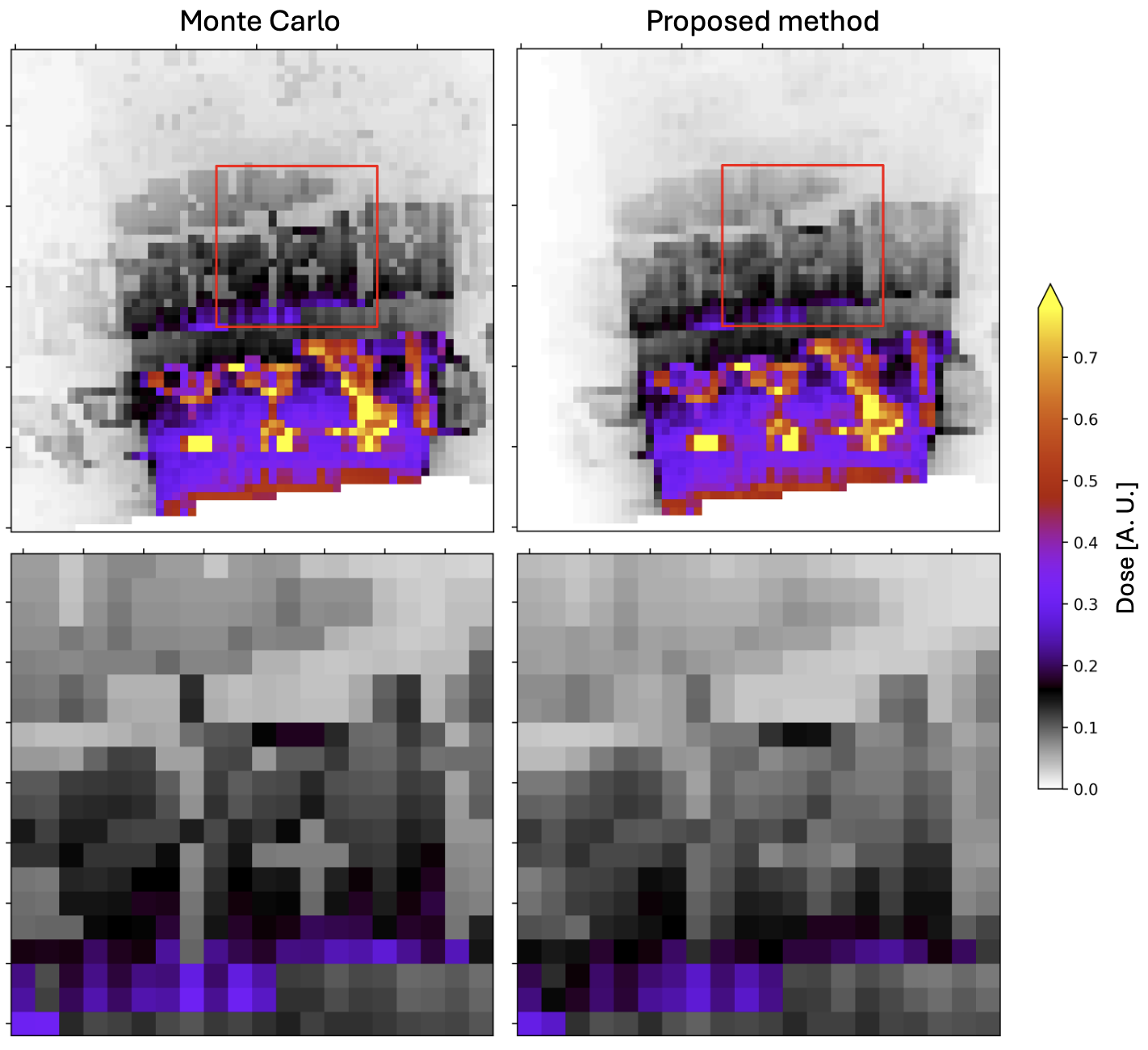}
	\caption{Dose maps for a sagittal slice (first row), with the ground truth (Monte Carlo Simulation, MCS) in the left column and the proposed method for 45,000 beamlets in the right column. The second row shows a zoomed-in view of the dose maps (as defined by the red rectangle).}
	\label{fig:X-Raydetail}
\end{figure}

Additionally, as observed in Figure~\ref{fig:X-RayConeBeam}, the noise generated by the proposed method differs from that of the Monte Carlo simulation (MCS). Specifically, the dose map tends to appear blurred compared to the ground truth, due to the convolutional nature of the deep learning approach. This effect is particularly noticeable in the vertebral column (see the right column of Figure~\ref{fig:X-Raydetail}).

The different metrics, MAE and ADE, were reported in Table~\ref{tab:X-RayConeBeam} for 9,000 and 45,000 beamlets. As expected, the metrics within the beam and for the larger number of beamlets were superior, with mean errors of 9.8\% and 4.9\% for 9,000 and 45,000 beamlets, respectively. As discussed, the errors relative to the Monte Carlo simulation (MCS) outside the beam were larger, at 15\% and 12\% for 9,000 and 45,000 beamlets, respectively. In this case, there was no significant improvement between 9,000 and 45,000 beamlets because the bias arose from the fixed patch size rather than the number of beamlets. The dose accuracy within the beam (~5\%) remained within clinically acceptable limits, particularly in medical imaging, where the primary objective was radiation protection and dose optimization.

\begin{table}
	\caption{Comparison between the proposed method and the MCS}\label{tab:X-RayConeBeam}%
	\begin{tabular}{@{}llll@{}}
		\toprule
		 & & 9,000 beamlets & 45,000 beamlets\\
		\midrule
		Within the beam  & MAE & 11.1$\times10^{-3}$ & 7.8$\times10^{-3}$ \\
		                              & ADE & 9.8\% & 4.9\% \\
		Outside the beam & MAE & 7.1$\times10^{-3}$  & 6.1$\times10^{-3}$ \\
		                                   & ADE & 15\% & 12\% \\
		\botrule
	\end{tabular}
\end{table}

Regarding computation times, obtaining the ground truth via Monte Carlo simulation with $5 \times 10^8$ particles required approximately 10 hours. Using the same system and using batches of 30 beamlets on the GPU (the optimal value), the network predicted the dose in 4 ms per beamlet. Additionally, the processing time for the patch (extraction and rotation) was around 2 ms per beamlet (using 6 CPU threads). For a dose map of 45,000 beamlets, the total time to obtain the map was 270 seconds (4.5 minutes). This represented a computational speed improvement of approximately 130 times compared to the Monte Carlo simulation.

\section{Discussions}
In this work, a hybrid method combining beamlet random sampling and deep learning approaches was explored for dose estimation in X-ray imaging. The aim was to improve 3D dose calculations using deep learning, but with a more generic solution that predicting dose maps for a given X-ray system. The versatility of the proposed solution allows for the simulation of multiple X-ray systems without the need to retrain the deep learning model with new beam specificities. The same trained model, learned using different anatomical sites, is capable of calculating the 3D dose within the patient for helical CT, cone-beam CT, fan-beam CT, or any collimated beam shape.

The dose prediction from the network has shown a relative absorbed dose error of 1.2$\pm$3.87\% compare to MCS. For a full example in cone beam CT, the proposed method has shown a faster convergence compare to standard Monte Carlo simulation. Although the results have shown good agreement between the beamlet approach and the ground truth from Monte Carlo simulation, with a relative absorbed dose error within the beam of around 5\%, some artifacts have appeared on the close border of the X-ray beam. Further investigation is needed to determine whether these artifacts originate from the network prediction or the patch processing and extraction.

One more limitation of the proposed method was the difficulty to recover dose that are outside the X-ray primary beam. This correspond to the dose deposited to the patient from the scattered particle. This come from the nature of the method that use patch-based approach. The very small dose values outside the patch are then not considered. However those dose values are two order of magnitude smaller than the dose inside the X-ray beam. One solution to solve this issue will be to predict the dose of a beamlet using the entire CT image, and then considering far scattered particle. This will solve the missing scattering dose but will slightly increase the computation time since the network will need to predict a large dose map. Regarding the current results, numerous improvement can be explored for improving the accuracy. This first study was a proof-of-concept showing promising result for solution that mix sampling approach and deep-learning based dose calculation for more versatile MCS. 

Although the proposed method was faster to converge compare to MCS, more improvement is possible regarding the running time for the patch extraction. This include the rotation of the oriented patch into axis aligned image, because the network has learned image in this frame, and the reverse operation that consist to rotate back the predicted dose into the final dose map. Two possible way can be used to improve the computation time. One consisting of developing fast image rotation using also deep learning or GPU programming. Another method will consist to directly learn an oriented patch image by the network with a specific architecture or using an image representation which is invariant in rotation.

\section{Conclusion}

A versatile approach that combines beamlet sampling and deep learning dose calculation was explored. The proposed idea was decomposing any x-ray beam into elemental small beams name beamlet. Then by using a sampling approach it is possible to simulate any beam geometry and energy. In this case only a neural network that predict the dose of a beamlet within patient CT was required. This allows multiple X-ray system simulations without retraining the model with new beam specificity. Results have shown that the dose engine based on deep learning have leaded a relative dose error about 1.2$\pm$3.87\% compare to the reference dose. For a more realistic simulation, in cone beam CT, dose results have shown a relative error within the beam of 5\% compare to a full MCS. The convergence of the proposed method was faster compare to MCS, with a speed up of x130 for equivalent dose results. This method is promising and required more investigation, especially to reduce artefact on the beam border and also to speed up image processing required to read and write ptach within the CT and dose map.

\backmatter

\bmhead{Acknowledgements}
This work was partially funded by the French National Research Agency through the MoCaMed project (ANR-20-CE45-0025).

\section*{Declarations}

\begin{itemize}
\item Funding: This work was partially funded by the French National Research Agency through the MoCaMed project (ANR-20-CE45-0025).
\item Conflict of interest/Competing interests: 
\begin{itemize}
	\item Maxime Rousselot declare that they have no known competing financial interests or personal relationships that could have appeared to influence the work reported in this paper.
	\item Jing Zhang declare that they have no known competing financial interests or personal relationships that could have appeared to influence the work reported in this paper.
	\item Didier Benoit declare that they have no known competing financial interests or personal relationships that could have appeared to influence the work reported in this paper.
	\item Chi-Hieu Pham declare that they have no known competing financial interests or personal relationships that could have appeared to influence the work reported in this paper.
	\item Julien Bert declare that they have no known competing financial interests or personal relationships that could have appeared to influence the work reported in this paper.
\end{itemize}
\item Ethics approval and consent to participate: Clinical data approval was obtained from the ethics committee of University Hospital of Brest. The procedures used in this study adhere to the tenets of the Declaration of Helsinki.
\item Consent for publication: All authors are consent for publication
\item Data availability: Not applicable
\item Materials availability: Not applicable
\item Code availability: Not applicable 
\item Author contribution: All authors contributed to the study conception and design.
\end{itemize}


\bibliography{photon_dose_biblio}

\end{document}